# Intermediate phase in the oxidative hydrothermal synthesis of potassium jarosite, a model *kagomé* antiferromagnet

W. Bisson [1], A. S. Wills[2, 3*],

[1]Davy Faraday Research Laboratory, The Royal Institution of Great Britain, 21 Albemarle Street, London, W1S 4BS
[2]Department of Chemistry, UCL, Christopher Ingold Laboratories, 20 Gordon Street, London, WC1H 0AJ
[3] The London Centre for Nanotechnology, 17-19 Gordon Street, London, WC1H 0AJ
[*] a.s.wills@ucl.ac.uk

**Keywords**: hydrothermal, jarosite, hydronium, kagomé, spin glass, frustration

**Abstract.** The jarosite family of minerals contain antiferromagnetically coupled $Fe^{3+}$ ions that make up the *kagomé* network. This geometric arrangement of the $Fe^{3+}$ ions causes magnetic frustration that results in exotic electronic ground states, *e.g.* spin glasses and spin liquids. Synthesic research into jarosites has focused on producing near perfect stoichiometry to eliminate possible magnetic disorder. An new oxidative synthesis method has been developed for the potassium, sodium, rubidium and ammonium jarosites that leads to high Fe coverage. We show through the identification of a meta-stable intermediate, using powder X-ray diffraction, how near perfect Fe coverage arises using this method. Understanding this new mechanism for jarosite formation suggests that is it possible to synthesis hydronium jarosite – an unconventional spin glass – with a very high Fe coverage.

## Introduction

The jarosite family of minerals are extensively studied in a wide range of research areas, from mining to magnetism [1-3]. The flexibility of the crystal structure allows the mineral to absorb many other elements and alter its stoichiometry, in turn influencing their magnetic and chemical properties. Knowledge of jarosite formation is crucial to understanding and manipulating the unstoichiometric nature and hence their chemical and magnetic properties.
Jarosites occur naturally through the weathering process of Pyrite, FeS [4]: high concentrations of $Fe^{2+}$ released from the dissolution of FeS decreases the pH of the local water area and through either microbial action or dissolved oxygen oxidises $Fe^{2+}$ to $Fe^{3+}$[5]. Around these $Fe^{3+}$ centres chains of hydroxysulphates attach themselves to form the characteristic iron flocs [6]. Extremely low values of pH and the incorporation of a suitably sized cation will cause jarosite precipitation and they have the following formula: $A_{1-x}(H_3O)_xFe_{3-y}(SO_4)_2(OH)_{6-3y}(H_2O)_{3y}$, where A=$K^+$, $Na^+$, $Ag^+$, $Rb^+$, $NH_4^+$, $H_3O^+$, $Pb^{2+}$, $Tl^{2+}$ [1]. Charge balancing requires that replacement of an A-site with a H$_2$O unit or protonation of the bridging OH unit be compensated by Fe vacancies [1]. This has important consequences for the magnetism as the Fe$^{3+}$ ions make up a *kagomé* network, figure 2b. Antiferromagnets with this geometry are of particular interest to fundamental magnetism as they are 'frustrated'. Hydronium jarosite has attracted much interest because it displays unconventional spin glass dynamics whereas all other jarosites order into a Néel state [2,3,7]. To understand these differences the role of the A-site in influencing the level of Fe vacancies must be determined.

In this article we demonstrate how a previously unobserved intermediate phase leads to the high Fe coverage observed in the oxidative method of jarosite synthesis [8].

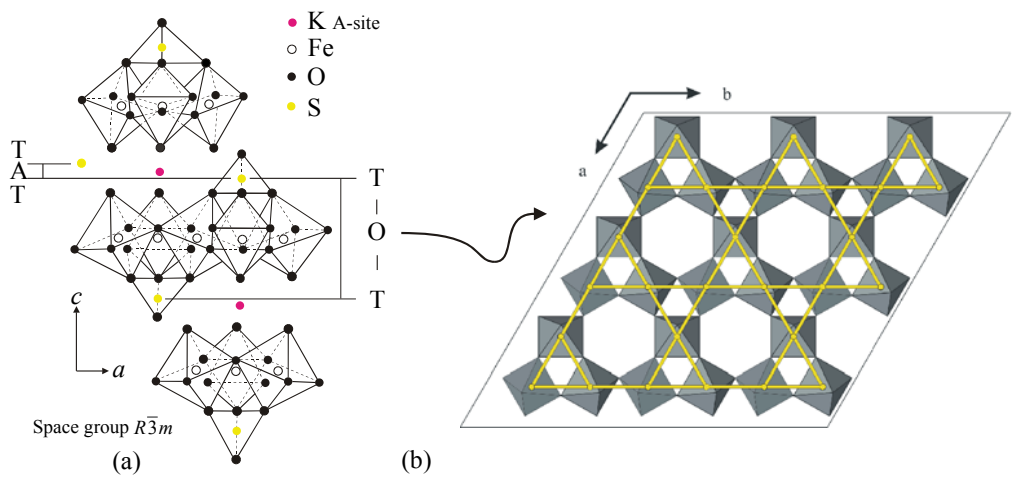

*Figure 1(a). A polyhedral representation of the jarosite structure. The iron coordination octahedra make up the kagomé networks as emphasised in 1(b). The iron coordination octahedra are capped by sulphate tetrahedra and the kagomé planes are separated by the A-site.*

## Crystal Structure

Jarosites have been intensely researched in the area of frustrated magnetism because the crystal structure contains the best known example of a *kagomé* network, a 2-dimensional lattice constructed Fe$^{3+}$ ions that link to form vertex sharing triangles. This geometry prevents antiferromagnetically coupled ions from satisfying the pair-wise interactions and leads to exotic electronic groundstates, such as spin glasses and spin liquids [2].

The jarosite structure is best described in the space group *R-3m* and has lattice parameters $a$~7.3 Å, $c$~17 Å. The *kagomé* plane is made up of iron coordination octahedra, as shown in figure 1b, and the Fe octahedra are capped above and below by sulphate tetrahedra. The 12 coordinate A-site, shown in figure 1a, separate these units. The A-site cation greatly influences the structure and the magnetism: all non-hydronium jarosites undergo a transition to a long ranged Néel state unlike hydronium jarosite which instead displays unconventional spin glass like properties and has been termed a topological or *kagomé* spin glass [7].

# Chemistry and synthesis

All jarosites are formed in highly acidic conditions pH<2 and non-hydronium A-site cations can be formed at temperatures below 100°C [1]. Hydronium jarosite, unique for its differing magnetic properties can only be synthesised under hydrothermal conditions [1] – supercritical water heated under pressure. Research has focused on these hydrothermal techniques because it improves the stoichiometry and increases crystal size for all the jarosites

A new hydrothermal oxidative process, similar to that used for zeolite synthesis [9] claims to achieve 100% Fe coverage and greater crystal size (>50μm across a face). This improvement to crystal quality has enabled greater insight into the crystal structure and magnetic characteristics. The drawback is that hydronium jarosite, arguably the most important jarosite for magnetism, cannot be synthesised via this method. This suggests there are at least two different reaction mechanisms for jarosite formation.

Synthesises of potassium jarosite used 38ml capacity Pyrex pressure tubes (Ace Glass Inc) with PTFE screw tops with a Viton O-ring seated inside the tube. 0.33g of Fe wire cut into pieces (2mm diameter, 99.9% purity) and 2.44g of $K_2SO_4$ were added to 25ml of $H_2O$ and 1.1ml of $H_2SO_4$, giving a filling of ~66%. The reaction was heated at 170°C for 2-3 days. A Pyrex pressure tube allows visual inspection, figure 2a, of the reaction and better control of the hydrothermal conditions which is not possible with a PTFE lined steel bomb.

A wide variation of hydronium jarosites were synthesised from MeOH/$H_2O$ solvent mixes [10]. 2g of $Fe_2(SO_4)_3 \cdot 5H_2O$ was dissolved into 15ml of MeOH/$H_2O$ mix. The mixtures ranged from 0%-90% MeOH. The samples were heated for 21 hours in a 23ml Pyrex pressure tube at various temperatures from 120-150°C

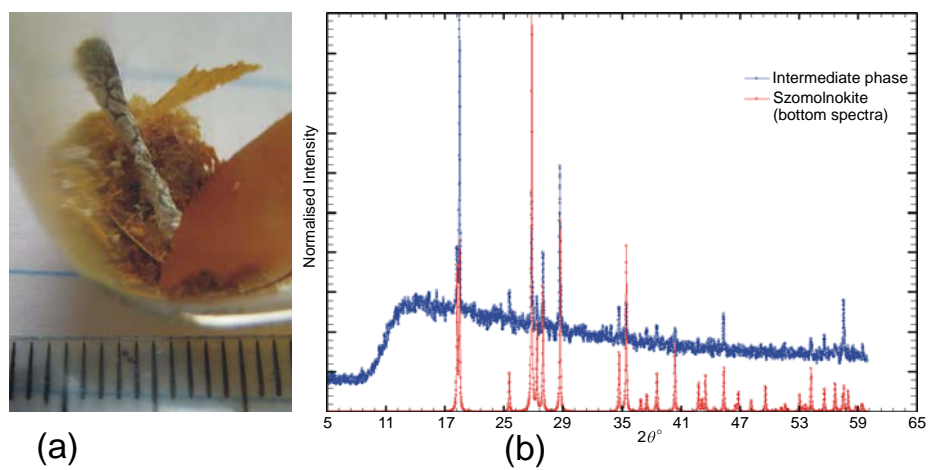

*Figure 2(a). Photograph of Szomolnokite (white) growing on the surface of the Fe wire. The image captured through a Pyrex tube also shows the dark ocherous colour of potassium jarosite surrounding the Fe wire. (b) The diffraction spectrum of the intermediate phase (top spectra) matched with that of synthetic Szomolnokite - $Fe^{2+}(SO_4) \cdot H_2O$ (bottom spectra) calculated with FullProf from the crystallographic data given in Table 1.*

# Szomolnokite FeSO$_4$.H$_2$O – meta-stable intermediate phase

The use of Pyrex tubes enabled the discovery an intermediate phase growing on the surface of the Fe wire, figure 2a, prior to potassium jarosite formation. The reaction was quenched in cold water and the intermediate was retrieved, and prepared for powder diffraction from acetone slurry. X-ray diffraction data were collected in Bragg-Bretano geometry on a low background silica plate using a Bruker D8 diffractometer with CuK$\alpha_1$ radiation and a PSD detector with a step size, $\Delta(2\theta)$=0.073°, 0.6s/ $\Delta(2\theta)$ and rotated at 15 rpm. Pattern matching using the software EVA showed it to be Szomolnokite – Fe$^{2+}$SO$_4$.H$_2$O. The data were only suitable for pattern matching due to the increased background from the Fe fluorescence. No contributions from jarosite phases were observed. Table 1 lists the crystallographic details.

*Table 1. The crystallographic details for Szomolnokite: Space group C2/c. Lattice parameters a=7.078 Å, b=7.549Å, c= 7.773 Å, α=90°, β=118.65°, γ=90° [11]*

| Atom | Wykoff site | x | y | z |
| --- | --- | --- | --- | --- |
| Fe | 4 b | 0 | 0.5 | 0 |
| S | 4 e | 0 | 0.15307 | ¼ |
| O1 | 6 f | 0.1697 | 0.0429 | 0.3985 |
| O2 | 6 f | 0.0956 | 0.2683 | 0.1560 |
| O3 | 4 e | 0 | 0.6444 | ¼ |
| H | 6 f | 0.108 | 0.709 | 0.315 |

# Mechanism for jarosite formation

Jarosite formation from the dissolution of Pyrite usually occurs via chains of bridging hydroxysulphates groups complexing the Fe$^{3+}$ ion. Continued jarosite formation decreases the pH, which promotes further jarosite formation:

FeS$_2$ + 3½O$_2$ + H$_2$O → Fe$^{2+}$ + 2SO$_4^{2-}$ + 2H$^+$ (dissolution of FeS)
Fe$^{2+}$ + H$_2$SO$_4$ + 1/2O$_2$ → Fe$^{3+}$ + SO$_4^{2-}$ + H$_2$O (oxidation to form ferric ions)
Fe$^{3+}$ + 5H$_2$O + SO$_4^{2-}$ → [Fe(H$_2$O)$_5$(SO$_4$)]$^+$ (jarosite precursor [5], stabilised at pH<2)
3[Fe(H$_2$O)$_5$(SO$_4$)]$^+$ + ½K$_2$SO$_4$ + ½H$_2$SO$_4$ → KFe$_3$(SO$_4$)$_2$(OH)$_6$ + 35H$^+$ + 2SO$_4^{2-}$

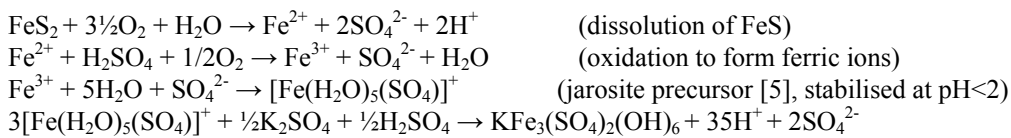

There are two key components that enable a reaction mechanism for the oxidative process to be derived. Firstly, hydrogen gas is evolved during this reaction. Secondly Szomolnokite does not form in the absence of high concentrations of an A-site sulphate. The reaction requires high pressure and temperature and a very low pH:

Fe + H$_2$SO$_4$ + ½O$_2$ + H$_2$O → Fe(SO$_4$).H$_2$O + H$_2$O
6Fe(SO$_4$).H$_2$O + K$_2$SO$_4$ + 6H$_2$O → 2KFe$_3$(SO$_4$)$_2$(OH)$_6$ + 3H$_2$ + 3H$_2$SO$_4$

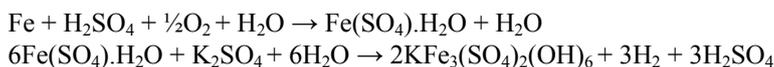

Structurally, jarosite shares many features, and in this case its origins, with Szomolnokite. The resemblance can be seen in the Fe coordination shown in figure 3. The Fe-O(S) distance within potassium jarosite is 2.076 Å, slightly longer than in Szomolnokite, at 2.053 Å. The Fe-O(H) distance, bridging the Fe centres in jarosite, is about 1.9806 Å, in Szomolnokite the Fe-O(H$_2$) distance is much longer at 2.228 Å. Oxidation of the Fe$^{2+}$ to Fe$^{3+}$ in Szomolnokite hydrolyses the H$_2$O molecules and draws the OH units closer to the Fe$^{3+}$ centres to allow bridging and subsequent formation of the Fe$^{3+}$ octahedrons that make up the *kagomé* plane.

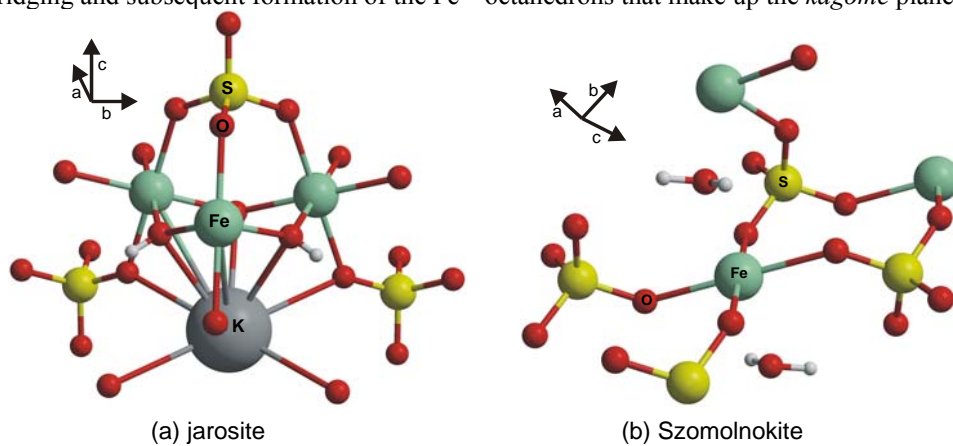

(a) jarosite      (b) Szomolnokite

*Figure 3(a). Structural representation of the jarosite structure. The sulphate groups cap above and below the Fe octahedra sitting along the 3-fold axis. Each of these subunits is separated by the A-site, K$^+$ in the diagram. (b) Szomolnokite has a more open structure; the distortion from rhombohedral is only slight: β ~ 120° and the a and b lattice parameters are similar to the a parameter in jarosite.*

What is the role of the A-site cation in jarosite formation and why does hydronium jarosite fail to form via the Szomolnokite intermediate? Firstly hydronium jarosite is rarely found in nature and can only readily be synthesized under hydrothermal conditions because of the need to stabilise the hydronium ion. Our hydronium jarosite synthesis have shown larger yields of hydronium jarosite are obtained with increasing MeOH concentration, and may result from the stability of the hydronium ion. A possible explanation for this is the reduction in hydration number, $h$, the number of H$_2$O molecules needed to solvate H$_3$O$^+$ when using MeOH/H$_2$O mixtures as a solvent. This is evidenced by lower values of the initial reaction pH. We find that too large a concentration of MeOH results in the formation of amorphous Fe oxyhydroxylsulphates with very high concentrations producing Mikasaite – Fe$_2$(SO$_4$)$_3$ (space group *R*-3) with no hydronium jarosite formed.

The hydration number, $h$, for K$^+$, Na$^+$, Rb$^+$ and NH$_4^+$ in solution, $h$=1.8±0.5, 3.9±0.5, 1.8±0.3 and 1.8±0.5 respectively [12], is significantly lower than H$_3$O$^+$, $h$=6.7±0.7 (H$^+$) [12]. MeOH for comparison is $h$=1±0.3 [12]. This suggests the incorporation of the non-H$_3$O$^+$ A-cation occurs either during the oxidation of Fe$^{2+}$ to Fe$^{3+}$ in Szomolnokite or immediately after the formation of the jarosite precursor. The lower mobility of the H$_3$O$^+$ cation highlights a different intermediate is involved in hydronium jarosite formation:

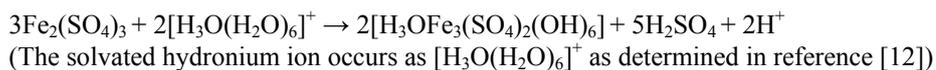
3Fe$_2$(SO$_4$)$_3$ + 2[H$_3$O(H$_2$O)$_6$]$^+$ → 2[H$_3$OFe$_3$(SO$_4$)$_2$(OH)$_6$] + 5H$_2$SO$_4$ + 2H$^+$
(The solvated hydronium ion occurs as [H$_3$O(H$_2$O)$_6$]$^+$ as determined in reference [12])

Minimal Fe vacancies are expected from the oxidative synthesis method because the charge neutrality of the Szomolnokite intermediate will reduce the need for charge compensation and the associated Fe vacancies. Formation of the jarosite can only be completed when the appropriate A-site sulphate reacts. It is therefore the high Fe occupation of Szomolnokite that keeps to a minimum the unstoichiometry of the resultant phase. Similarly, negligible Fe vacancies within hydronium jarosite formed by conventional hydrothermal synthesis as are expected: the jarosite precipitation occurs via the likely intermediate, Mikasaite, where the Fe ions are fully oxidised. Again the intermediate is charge balanced which leads to near 100% occupancy of $H_3O^+$ ions in the A-site of the jarosite and again very high Fe occupancy. Fe vacancies may, however, arise in the $MeOH/H_2O$ preparations where other precursors lead to the formation of amorphous Fe oxyhydroxylsulphates. Increasing concentration of MeOH produces a greater number of alternative nucleation sites, and through the process of Oswald ripening, pushes the precipitation away from the thermodynamically stable jarosite product to amorphous Fe oxyhydroxylsulphates phases and poor quality jarosites.

## Conclusions

The discovery of Szomolnokite as an intermediate phase in the oxidative synthesis procedure has revealed more about jarosite formation mechanisms and the ability to minimise structural defects for all jarosites. With this knowledge of jarosite formation we suggest that an intermediate, solvated Mikasaite, is involved in hydronium jarosite formation and leads to very high Fe coverage. These intermediates and the role of the A-site cation which ultimately directs the final structural composition, all act to minimise Fe deficiencies. The ability to achieve near perfect stoichiometry for all jarosites with this mechanism can further research into their crystal chemistry and their frustrated magnetic properties.